\documentclass[aip, cha,reprint,amsmath,amsfonts]{revtex4-1}

\usepackage{amssymb}
\usepackage{amsmath}
\usepackage{bm}
\usepackage{paralist}
\usepackage{graphicx}
\usepackage{enumerate}
\usepackage{tikz}
\usepackage{graphicx}
\usepackage{verbatim}
\usepackage{etoolbox}
\usepackage{nicematrix}

\usetikzlibrary{matrix,positioning,fit}

\usepackage{hyperref}
\hypersetup{
	colorlinks,
	citecolor=blue,
	filecolor=blue,
	linkcolor=blue,
	urlcolor=blue,
	pdfproducer={}
}

\let\bbordermatrix\bordermatrix
\patchcmd{\bbordermatrix}{8.75}{4.75}{}{}
\patchcmd{\bbordermatrix}{\left(}{\left[}{}{}
\patchcmd{\bbordermatrix}{\right)}{\right]}{}{}

\newlength{\Mylen}

\usepackage{ragged2e}
\usepackage{caption}
\usepackage{ragged2e}
\DeclareCaptionJustification{justified}{\justifying}
\usepackage{soul}	

\usepackage [autostyle, english = american]{csquotes}
\MakeOuterQuote{"}

\usepackage[normalem]{ulem}

\usepackage[percent]{overpic}

\usepackage{xcolor}

\usepackage{cprotect}

\begin{document} 

\title{Measures of Chaotic Advection in Simulations of Active Nematics}

\author{Md Mainul Hasan Sabbir}
\email{msabbir@ucmerced.edu}
 \affiliation{Physics Department, University of California, Merced, CA 95344, USA.}

 \author{Brandon Klein}
 \affiliation{Department of Physics and Astronomy, Johns Hopkins University, Baltimore, MD 21218, USA.}

 \author{Daniel A. Beller}
 \affiliation{Department of Physics and Astronomy, Johns Hopkins University, Baltimore, MD 21218, USA.}

 \author{Kevin A. Mitchell}
\email{kmitchell@ucmerced.edu}
 \affiliation{Physics Department, University of California, Merced, CA 95344, USA.}

\date{\today}
	
\begin{abstract}

Active nematics are non-equilibrium fluids composed of rod-like self-propelled units that collectively generate large-scale coherent flows. Here, we focus on a canonical experimental system: an active nematic fluid in $2D$ driven by adenosine triphosphate (ATP), composed of densely packed, extended microtubule bundles cross-linked by kinesin motors. An intriguing feature of this system is the creation and annihilation of topological defects with topological charge $\pm1/2$, due to the fracturing of the material. Experiments confirm that the positive ($+1/2$) defects serve as \emph{virtual stirring rods} that move around each other in a complex braiding pattern. This collective braiding motion of positive defects stretches and folds the fluid itself, i.e., produces macroscale chaotic advection. The degree of self-mixing due to the chaotic advection can be measured using \textit{topological entropy} and the \textit{Lyapunov exponent}. Our goal is to determine whether continuum models of microtubule-based active nematic fluid can reproduce these measurements. To this end, we use two continuum models: the traditional Beris-Edwards (BE) model and the more recently developed Beris-Edwards model with enhanced nematic locking (BENL). The difference between the two models is the adoption of the \emph{nematic locking principle}, which states that an individual microtubule bundle cannot rotate independently of its neighbors due to steric interactions among elongated dense microtubule bundles. This principle holds in the BENL model except in small localized areas of the material domain where the material fractures, specifically near the creation and annihilation of topological defects. We employ several numerical methods to estimate measures of chaotic advection using both models. Our study shows that the BENL model more accurately reproduces experimental measures of self-mixing driven by chaotic advection in microtubule-based active nematic fluids.

\end{abstract}

\maketitle

\section{Introduction}

Active matter is a class of non-equilibrium systems composed of self-driven, interacting agents that self-organize to produce large-scale spontaneous collective motion~\cite{Das2020, Bowick2022, Joanny2023}. The defining feature of active matter compared to other non-equilibrium systems is that each individual agent consumes locally available energy and converts it to drive the spontaneous motion rather than using an external source of energy~\cite{Marchetti2013}. Examples of active matter span a wide range of length scales, from flocks of birds~\cite{Toner1995} to bacterial suspensions~\cite{Aranson2022}. Active matter systems exhibit a wealth of non-equilibrium collective dynamical behaviors with emergent large-scale structures because of the interplay between local energy consumption and interactions among the constituent agents --- an inherent property of living systems. Thus, studying active matter provides a unique opportunity to develop theories of the dynamics and function of living systems and to use the acquired knowledge of their dynamics to design next-generation bio-inspired materials~\cite{Gompper2025}.

Active fluids are a class of active matter that exhibit large-scale coherent flow~\cite{Saintillan2018}. Notable examples of active fluids are swarms of bacteria~\cite{Sokolov2007, Dunkel2013}, epithelial cell layers~\cite{Saw2017, Perez2019, Armengol2023} or mixtures of biological polymers and molecular motors~\cite{Sanchez2012, Nitin2018, Duclos2020}. In this article, we study microtubule-based active nematics, a bioengineered active fluid~\cite{Sanchez2012, Keber2014, Guillamat2018, Thijssen2021}. This fluid is an aqueous mixture of a densely packed and orientationally ordered $2D$ layer of extended microtubule filaments cross-linked by kinesin motors. Consuming energy via hydrolysis of ATP, kinesin motors slide the microtubules relative to one another. These collective movements generate extensile stresses that deform the microtubule bundles in regions of high curvature in the material domain. When the microtubule bundle curvature exceeds a certain limit, the bundles fracture, creating localized regions of zero density where the orientational order breaks down, known as topological defects.

\begin{figure}
 \centering
\includegraphics[width =  \columnwidth]{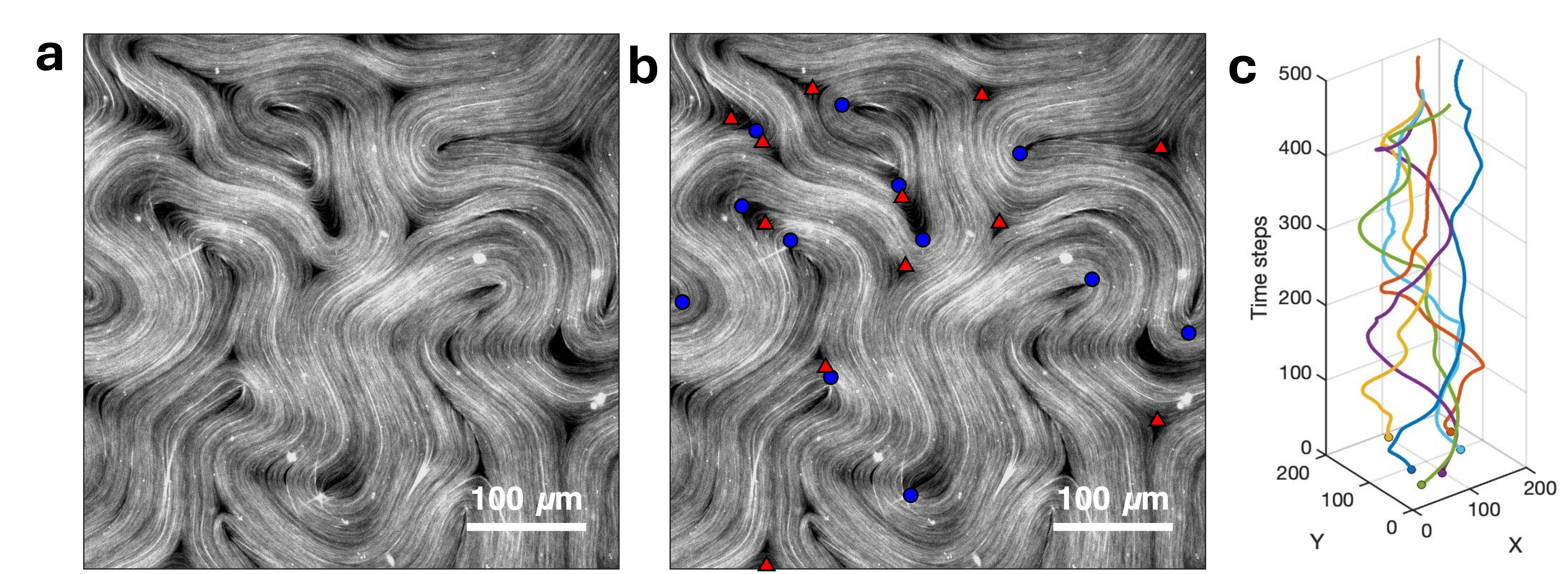}
\caption{ a) Fluorescence microscope image of microtubule-based active nematic fluid (Experimental data courtesy: Hirst Lab, UC Merced)~\cite{Tan2019}. b) Topological defects are shown as blue circles (positive $+1/2$ defects) and red triangles (negative $-1/2$ defects). c) An example of a space-time diagram of six $+1/2$ defect trajectories (represented by six different colors) using simulated data. (See Sec.~\ref{sec:etec} for more details.)}
\label{fig:exp}
\end{figure}

The orientational order formed due to the local alignment of extended microtubule bundles appears as striation patterns in fluorescence microscopy images (Fig.~\ref{fig:exp}a). The orientational order breaks down in the vicinity of topological defects, i.e., singularities in the orientational order. These defects are mobile, created in pairs with topological charges of $\pm 1/2$ (Fig.~\ref{fig:exp}b). These topological defects move around one another in an entangled braiding pattern to produce macroscale chaotic advection~\cite{Aref2017, Aref2020}, i.e., exponential separation of nearby fluid patches caused by stretching and folding. Thus, the fluid self-mixes through continuous, self-sustained stretching and folding motion. This dynamic steady state is commonly referred to as \emph{active turbulence}~\cite{Thampi2016, Alert2022}, which occurs despite an extremely low Reynolds number(the ratio between inertial and viscous forces). Active turbulence is also seen in living systems~\cite{Wensink2012, Baruah2026}. Therefore, the microtubule-based active nematic fluid serves as an experimental model system for understanding mixing and transport driven by chaotic advection in living systems~\cite{Wensink2012, Baruah2026}, or, more broadly, mixing at low Reynolds numbers~\cite{Ober2015}.

The mixing driven by chaotic advection can be measured using the topological entropy $h$ and the Lyapunov exponent $\lambda$~\cite{Hassan2017}. The topological entropy measures the exponential stretching rate of the material line. We can imagine measuring topological entropy by putting a line of colored dye in the fluid and tracking its growth as it gets stretched and folded. Another perspective on measuring topological entropy is based on braid theory~\cite{Thiffeault2005, thiffeault2022book}. The particle trajectories form \emph{braids} in space-time. Braids are algebraic objects that store information about how trajectories exchange positions with respect to each other. Plotting these trajectories in a three-dimensional graph, with time being the vertical axis, we can get a \emph{spaghetti plot}, i.e., a space-time diagram of trajectories (Fig.~\ref{fig:exp}c). A lower bound of the topological entropy of the flow can be measured from the degree of entanglement of these trajectories~\cite{Thiffeault2010, Budisic2015}.

The Lyapunov exponent $\lambda$ is another measure of mixing that quantifies the rate at which nearby passive tracers separate from each other. The measurement of the Lyapunov exponent $\lambda$ is sensitive to initial conditions, i.e., the initial position of the passive tracer trajectories and the smooth structure of the underlying flow. The connection between the two measures of chaotic advection described above is a well-established result in dynamical systems -- the topological entropy is the upper bound of the metric entropy (also known as measure-theoretic or Kolmogorov–Sinai entropy)~\cite{Goodman1971, Eckmann1985} for smooth flows. In two-dimensional flows, the Lyapunov exponent equals the metric entropy.  Thus, the topological entropy is an upper bound to the Lyapunov exponent~\cite{Pesin1977, Eckmann1985}.


Tan et al.~\cite{Tan2019} first performed an experimental measurement of chaotic advection to quantify the self-mixing of microtubule-based active nematic fluid. The authors measured the topological entropy $h$ using defect trajectories, grouping them into three categories: all defect trajectories, only positive defect trajectories, and only negative defect trajectories. They estimated the topological entropy of the active nematic flow using the braiding pattern of each group of defect trajectories. Their analysis showed that the positive defects act like \emph{virtual stirring rods} that self-mix the material. Using ergodic conditions, i.e., assuming that the fluid is sufficiently well mixed, Ref.~\onlinecite{Tan2019} also estimated the Lyapunov exponent of the flow. They observed that the different measures of topological entropy are slightly larger than the Lyapunov exponent. These results draw a connection between mixing due to global defect topology and mixing due to local fluid deformation in an active nematic fluid. Later, Memarian et al.~\cite{Memarian2024} experimentally demonstrated that cardioid-shaped confinement can lead to a maximally efficient mixing state where three positive defect trajectories form a braid like a figure-eight pattern that is topologically identical to a \emph{golden braid}. The spontaneous braiding of topological defects is also found to be crucial in the dynamics of protein signaling and force generation on the cell membrane~\cite{Liu2021}. However, a systematic study on measures of chaotic advection using theoretical models of active fluids is still in its infancy~\cite{Klein2026}.

Inspired by the above-mentioned experiments, in this article we ask the question: Can continuum models of two-dimensional microtubule-based active nematic fluids reproduce the experimental measures of chaotic advection, and if so, what aspects of the theoretical model are necessary to ensure this? To accomplish our goal, we first analyze the well-established Beris-Edwards (BE) model. However, the BE model has some limitations in modeling microtubule-based active nematic fluid. One notable limitation is the treatment of microtubules as individual, small entities rather than extended bundles. To address this, Mitchell et al.~\cite{Mitchell2025} modified the BE model based on a physical principle named the \emph{nematic locking principle}. This principle states that individual microtubule bundles cannot rotate independently of their neighboring bundles due to steric interactions, a condition that holds throughout the material domain except in localized regions of high curvature where the material fractures, e.g., at points of defect-pair creation and annihilation. In other words, the microtubule-based active nematic fluid is dominated by nematic locking throughout most of the fluid domain, where the rotations of the director field (nematic order) are \emph{locked} to the fluid velocity field. The modified Beris-Edwards model with enhanced nematic locking (BENL) ensures that fracturing occurs in localized areas rather than throughout the entire material domain. We employ several numerical methods to estimate the topological entropy and the Lyapunov exponent of the active nematic fluid. Then we compare our results from both continuum models with experimental measures of chaotic advection. We report that the BENL model aligns better with experimental observations of chaotic advection in the microtubule-based active nematic fluid than the BE model.

The article is organized as follows: Sect.~\ref{sec:models} describes the two-dimensional continuum models that we use in our study. In Sect.~\ref{sec:measures}, we describe methods for estimating measures of chaotic advection and apply them using simulated data from continuum models. We summarize the topological entropy analysis in Sect.~\ref{subsec:topEn} and the Lyapunov exponent analysis in Sect.~\ref{subsec:LE}. Lastly, we summarize all of our findings in Sect.~\ref{sec:end}.

\begin{figure}
 \centering
\includegraphics[width = 0.95  \columnwidth]{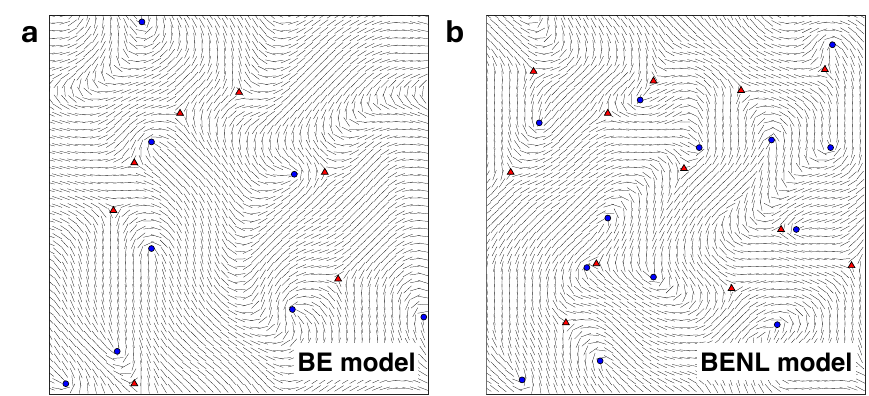}
\caption{A single snapshot of simulations of a) BE model, b) BENL model for active length scale $l_{a} =3$ in a $200 \times 200$ simulation domain with periodic boundary conditions. We use the same initial conditions and parameters for both models. The director field is represented by black rods. Positive $+1/2$ topological defects are blue circles and negative $-1/2$ topological defects are shown in red triangles. \label{fig:simSnaps} }
\end{figure}

\section{Continuum models of active nematics}
\label{sec:models}

\subsection{Beris-Edwards (BE) Model}
\label{sec:BE}

The continuum description of the dynamics of the microtubule-based active nematic is governed by two fields: the nematic order field $\mathsf{Q}$ and the fluid velocity $\mathbf{u}$. The nematic tensor $\mathsf{Q}$ (traceless and symmetric) is defined as 
\begin{equation}
  \mathsf{Q} = S  (\mathbf{n} \otimes \mathbf{n} - \mathsf{I}/2), \label{e1}
\end{equation}
where $S$ is the scalar order parameter that quantifies the degree of alignment, e.g., $S = 1$ corresponds to the rod-like particles being in perfect alignment along some direction, whereas $S = 0$ corresponds to randomly oriented particles. The director $\mathbf{n}$ represents the local average direction, with $\mathbf{n} = - \mathbf{n}$ to reflect the head-tail symmetry of the rod-like particles. 

An extensively used continuum model for the evolution of the $\mathsf{Q}$-tensor is the Beris-Edwards (BE) model~\cite{Beris1994}. In the BE model, the evolution of the $\mathsf{Q}$-tensor is modeled by the following equation:
\begin{align}
  \frac{D}{D t}  \mathsf{Q} &= \frac{\partial}{\partial t} Q \;+\; \mathbf{u} \cdot \nabla Q \notag \\
 &=\lambda_a S \mathsf{E} +  [\mathsf{Q}, \Omega ] -
2 \lambda_a \text{Tr} (\mathsf{Q} \mathsf{E} ) \mathsf{Q}  +
\frac{1}{\gamma} \mathsf{H},
\label{eq:BEq}
\end{align}
where $[,]$ is the commutator, $\mathsf{E} = (\nabla \mathbf{u} + \nabla \mathbf{u}^T)/2$ is the symmetric strain‐rate tensor and $\Omega = (\nabla \mathbf{u} - \nabla \mathbf{u}^T)/2$ is the antisymmetric vorticity tensor. $\mathsf{E}$ characterizes the deformation (stretching) and $\Omega$ characterizes the rotation  of the flow field $\mathbf{u}$. The flow alignment parameter $\lambda_a$ determines how the director field responds to the flow field (In the literature, this parameter is generally denoted by $\lambda$). Recent works have shown quantitatively that $\lambda_a=1$ best describes the microtubule-based active nematic fluid~\cite{Joshi2022, Matthew2023, Mitchell2024, Mitchell2025}. The first three terms in Eq.~(\ref{eq:BEq}) are nonlinear advection terms that originate from the passive advection of the microtubule bundles. The last term in Eq.~(\ref{eq:BEq}) describes the relaxation of the $\mathsf{Q}$-tensor towards the minimum of free energy where $\gamma$ is the rotational viscosity and $\mathsf{H}$ is the molecular tensor defined as
\begin{equation}
    \mathsf{H} = - \left(\frac{\delta F_{LdG}}{\delta \mathsf{Q}}
    - \frac{1}{2}\operatorname{Tr}\frac{\delta F_{LdG}}{\delta \mathsf{Q}} \right).
    \label{eq:h}
\end{equation}
Here, $\operatorname{Tr}$ is the trace, and $F_{LdG}$ is the Landau-de Gennes free energy
\begin{align}
F_{LdG} &= \int_D \left[ \frac{A}{2} \operatorname{Tr}(\mathsf{Q}^2) +
\frac{C}{4} \left(\operatorname{Tr}(\mathsf{Q}^2)\right)^2 \right]\, da \nonumber \\
&\quad + \int_D \frac{1}{2} K (\nabla_i Q_{jk})(\nabla_i Q_{jk})\, da,
\label{eq:fldg}
\end{align}
where $K$ is the elastic constant, and $D$ is the material domain. The first term in Eq.~(\ref{eq:fldg}) is the phase free energy that controls the isotropic to nematic transition. The last term in Eq.~(\ref{eq:fldg}) is the elastic energy that penalizes any spatial inhomogeneities in the nematic tensor field $\mathsf{Q}$. The molecular tensor $\mathsf{H}$ in Eq.~(\ref{eq:h}) can be rewritten using Eq.~(\ref{eq:fldg}) by decomposing it into a phase term and an elastic term as
\begin{align}
\mathsf{H} &= -\mathsf{Q}\left(A + C\,\operatorname{Tr}(\mathsf{Q}^2)\right) + \frac{1}{2} K \nabla^2 \mathsf{Q},
\label{e5}
\end{align}
where the first term of Eq.~(\ref{e5}) is the phase term, and the last term of Eq.~(\ref{e5}) is the elastic term. We choose $C = -2A>0$ so that the integrand of Eq.~(\ref{eq:fldg}) is a double well in $S$, with $S = 1$ the location of the bottom of the right well.

The second dynamic equation is for the fluid velocity field governed by the Navier-Stokes equation in the Stokes limit
\begin{equation}
  \nabla_j \Pi_{ij} = 0,
  \label{e6}
\end{equation}
where we assume uniform density $\rho = 1$. The stress tensor is
\begin{equation}
  \Pi = 2\eta \mathsf{E} - p \mathsf{I} +\Pi^E + \Pi^A.
  \label{eq:stress}
\end{equation}
The first two terms in Eq.~(\ref{eq:stress}) are the viscous damping, with viscosity $\eta$, and the pressure $p$. The elastic stress is
\begin{align}
  \Pi^E & = -\lambda_a \left( S \mathsf{H} -  2 \text{Tr}(\mathsf{QH}) \mathsf{Q}\right) +
  [\mathsf{Q},\mathsf{H}]  -K
\nabla Q_{ij} \otimes
\nabla Q_{ij}.
\label{e8}
\end{align}
Finally, the active stress responsible for the extensile driving of the system is given by
\begin{equation}
\Pi^A = -\zeta \mathsf{Q},
\label{eq:actstress}
\end{equation}
with activity parameter $\zeta > 0$ for extensile systems. The active stress in Eq.~(\ref{eq:actstress}) comes from considering the symmetry of the flow field generated by rod-like particles~\cite{Edwards2009}.

Two length scales can be defined for the bulk material: the active length scale $l_{a} = \sqrt{K/\zeta}$ and the nematic coherence length $l_{n} = \sqrt{K/C}$. $l_{a}$ is the length scale over which elastic and active stresses balance, and $l_{n}$ is the length scale that determines how quickly the nematic order drops near topological defects and can be taken as a measure of defect core size~\cite{Giomi2015}.  

\subsection{Beris-Edwards Model with Enhanced Nematic Locking (BENL)}

In the BE model, the material density is assumed to be uniform, although we observe density variations in the experimental images, as indicated by striation patterns. To circumvent the issue, the scalar order parameter $\mathsf{S}$ works as a \emph{proxy} for the density of the material such that $\mathsf{S}$ goes to zero at defect cores where the material density is low. The \emph{nematic locking principle} implies that microtubule bundles can rotate independently in localized areas of low density where $S \approx 0$, i.e., in the vicinity of the creation and annihilation of topological defects. Conversely, microtubule bundles cannot rotate individually where material density is high due to steric interactions among long, extended microtubule bundles. 

The main difference between the BE model and the BENL model is how each model relates to the \emph{nematic locking principle} in the material domain. The BE model violates the \emph{nematic locking principle} throughout the material domain, specifically via the elastic energy term of the molecular tensor $\mathsf{H}$~\cite{Mitchell2025}. This term is crucial for the creation and annihilation of defects. However, this term must be suppressed throughout the material domain except in areas of low density, i.e., $S \approx 0$, in order to obey the \emph{nematic locking principle}. To achieve that, Ref.~\onlinecite{Mitchell2025} decomposed $\mathsf{H}$, a $2 \times 2$ symmetric, traceless matrix, into a linear combination of the two basis matrices $\mathsf{Q}$ and $\mathsf{U = JQ}$, where $\mathsf{J}$ is the right-handed rotation by $\mathsf{\pi/2}$,
\begin{equation}
  \frac{1}{\gamma} \mathsf{H} = \frac{2}{\gamma S^2}[\text{Tr}(\mathsf{HQ}) \mathsf{Q} + \text{Tr}(\mathsf{HU}) \mathsf{U}].
  \label{e10}
 \end{equation}
Ref.~\onlinecite{Mitchell2025} noted that any term proportional to $\mathsf{U}$ violates the nematic locking principle. By modifying the prefactor of the $\mathsf{U}$ term in $\mathsf{H}$, (Eq.~(\ref{e10})), with a nonlinear switch function of $S$, the $\mathsf{U}$ term can be localized in areas where $S \approx 0$.
\begin{align}
 \frac{1}{\gamma} \mathsf{H}  & \rightarrow \frac{1}{\gamma} \frac{2}{S^2}[\text{Tr}(\mathsf{HQ}) \mathsf{Q} + e^{-S^2 /(2
     \sigma^2)} \text{Tr}(\mathsf{HU}) \mathsf{U}] \label{r14}\\ 
   &=
     \frac{1}{\gamma} \mathsf{H}+ \frac{1}{\gamma}\frac{2 (e^{-S^2 /(2
     \sigma^2)}-1)}{S^2} \text{Tr}(\mathsf{HU}) \mathsf{U}.
\label{e11}
\end{align}
So, the modified equation for the evolution of the nematic tensor $\mathsf{Q}$ for the BENL model can be written as follows:
\begin{align}
  \frac{D}{D t}  \mathsf{Q}
 &= S \mathsf{E} +  [\mathsf{Q}, \Omega ] -
2 \text{Tr} (\mathsf{Q} \mathsf{E} ) \mathsf{Q}  + \frac{1}{\gamma} \mathsf{H} \nonumber \\
&+ \frac{1}{\gamma}\frac{2 (e^{-S^2 /(2
     \sigma^2)}-1)}{S^2} \text{Tr}(\mathsf{HU}) \mathsf{U}.
     \label{eq:BENL}
\end{align}
The only difference between the BE model and the modified BENL model is the final term in Eq.~(\ref{eq:BENL}). The Navier-Stokes equation governing the fluid velocity remains the same as in the BE model for the BENL model. Full details of the BENL model can be found in Ref.~\onlinecite{Mitchell2025}. 

\subsection{Numerical Simulation Parameters}

We use simulated velocity and $\mathsf{Q}$-tensor fields to estimate measures of self-mixing driven by chaotic advection. In the numerical simulation of both models (BE and BENL), we use the following parameters: $\lambda_a = 1$, $\gamma = 5 \times 256$, $C = 256^2$, $K = 256^2$, $\eta = 2560$, and $\zeta = (256/3)^2$. Using these parameters, we can calculate the following derived quantities.  The Reynolds number is $\text{Re} = K/\eta^2 = 0.01$ (Stokes regime), the active length is $\ell_a = \sqrt{K/\zeta} = 3$, and the nematic coherence length is $\ell_n = \sqrt{K/C} = 1$. We use $\sigma = 0.2$ for the switch function in the BENL model. The simulation domain is $200 \times 200$ with periodic boundary conditions. We used an initial director field with randomly chosen orientations and an initial velocity $\mathbf{u} = 0$. In this article, we use the integration time unit that we define as $t = frames \times \Delta t_{sim}$, where $\Delta t_{sim} = 1.79 \times 10^{-5}$ is the simulation time of the model per frame. The numerical values of the topological entropy and the Lyapunov exponent are reported in inverse integration time units, i.e., $[\mathrm{itu}^{-1}]$.

\section{Measures of Chaotic Advection}

\label{sec:measures}

\subsection{Topological Entropy}


Topological entropy $h$ is one of the fundamental measures in dynamical systems that quantifies the exponential rate at which the number of distinguishable trajectories grows over time~\cite{Adler1965, Wilczak2025}. Hence, the topological entropy is a measure of the complexity of a dynamical system. The definition of the topological entropy for a continuous flow $f$ in a compact metric space is
\begin{equation}
    h(f)
  = \lim_{\varepsilon \to 0}
    \left\{
      \lim_{T \to \infty} \text{sup} \frac{1}{T} \log N(T,\varepsilon)
    \right\}, 
    \label{eq:htop}
\end{equation}
where $N(T,\varepsilon)$ is the number of distinguishable orbits of time duration $T$ and $\varepsilon$ is a finite resolution scale below which it is difficult to identify distinguishable trajectories~\cite{Wilczak2025}. For a chaotic flow, the number of distinguishable trajectories at a finite resolution grows exponentially with time. Topological entropy has been widely used in fluid dynamics to study mixing in two-dimensional fluid flows~\cite{Thiffeault2006, Finn2007, Thiffeault2010, Candelaresi2017}. 

 We estimate the topological entropy of active nematics using the simulated velocity $\mathbf{u}$ from both models (BE and BENL).  We employ trajectory-based Lagrangian methods. First, we use the line stretching algorithm, which is based on the geometric evolution of a material line (or curve) embedded in the flow. In this method, the material line is viewed as a continuum of passive tracers advected by the fluid velocity. We track the growth of the Euclidean length of the line segment as it deforms, stretches, and rotates over time. Next, we use the E-tec algorithm~\cite{Roberts2019, EtecCode}. E-tec takes the trajectories of passive tracers and uses them to form entangled braids. Then we calculate the topological entropy of the braid from these entangled trajectories~\cite{Thiffeault2010, Candelaresi2017}. In the following subsections, we introduce each method and use it to estimate the topological entropy of the flow.

\subsubsection{Line Stretching Algorithm}

\begin{figure*}[t]
 \centering
\includegraphics[width = 2.0 \columnwidth]{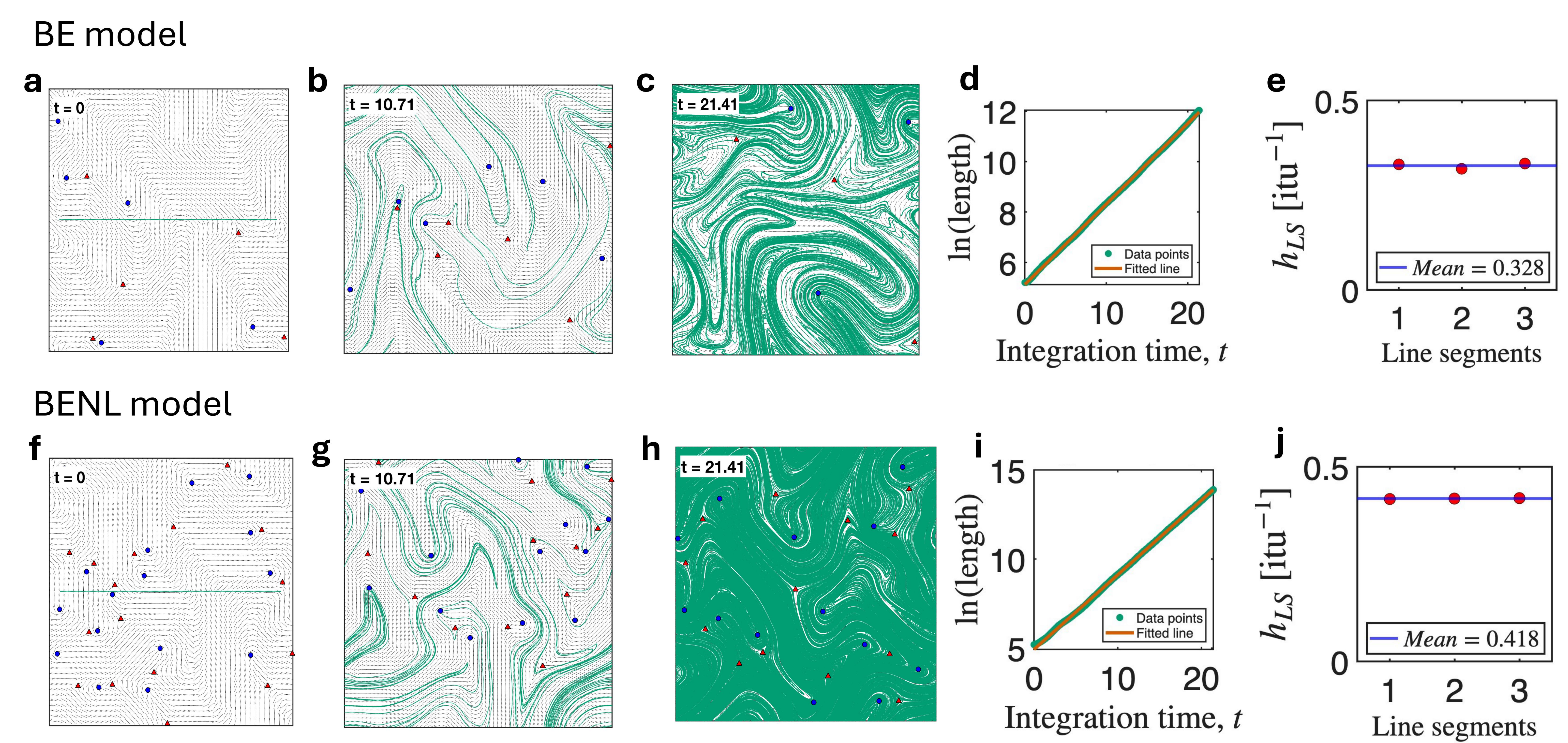}
\caption{Summary of the analysis using the line-stretching algorithm for both the BE and the BENL models. (a, f) An initial line segment with uniformly distributed tracers at $t=0$. (b, g) The advected curve at an intermediate integration time $t =10.71$. (c, h) The advected curve at the final integration time $t =21.41$. (d, i) Semi-log plot of the length of the line segment vs. integration time, $t$. (e, j) Red circles are the topological entropy $h_{LS}$ in inverse integration time units for three initial line segments, and the blue line represents the average over these three segments. \label{fig:LSana} }
\end{figure*}

The conceptually simplest way to estimate topological entropy $h$ for a two-dimensional flow is to use the rate of growth of a line segment~\cite{Newhouse1993}
which might, e.g., be a nematic contour line (integral curve of the director field). Let $L_0$ be the Euclidean length of a line segment at time $t= 0$. The initial line segment stretches and folds as it is advected by the flow. Then, at long times, the Euclidean length $L(t)$ of the advected curve as a function of time $t$ grows exponentially as
\begin{equation}
  L(t) = L_{0} e^{ht},
\end{equation}
where $h$ is the topological entropy and has units of inverse time. We implement the line stretching algorithm by initializing a line segment using uniformly distributed points, where each point represents a passive tracer in the fluid. Then we evolve the initial line segment forward in time using the fluid velocity. We record the length of the curve at each timestep. The exponential growth rate, i.e., the topological entropy, can be calculated from the slope of a semi-log plot of the Euclidean length of the line segment vs. time. Note that we introduce new passive tracers on the curve when the Euclidean distance between neighboring passive tracers goes beyond a certain limit to maintain the smoothness of the advected curve. This is critical because in chaotic flows, the length of the line segment grows exponentially fast. Therefore, the algorithm requires an exponentially growing number of trajectories to maintain sufficient point (tracer) density in the advected curve. Thus, this algorithm becomes exponentially expensive in time. 

Fig.~\ref{fig:LSana} shows the evolution of a line segment using the simulated flow fields from both the BE and BENL models.
We chose a horizontal line segment at time $t=0$ (Fig.~\ref{fig:LSana}a and ~\ref{fig:LSana}f). Then, using the underlying flow field, the line segment undergoes stretching and folding, and the length grows over time. The final advected line segments for both the BE and the BENL models are shown in Fig.~\ref{fig:LSana}c and Fig.~\ref{fig:LSana}h. We calculated the slope from the semi-log plot of length vs. integration time (Fig.~\ref{fig:LSana}d and \ref{fig:LSana}i). The value of the slope is an estimate of the topological entropy $h_{LS}$ of the flow. We use three different line segments to calculate the mean topological entropy $h_{LS}$ with units of inverse integration time and the standard error of the mean (supplementary Fig. S1). We obtain $h_{LS} =0.328 \pm 0.004$ for the BE model and $h_{LS} =0.418 \pm 0.001$ for the BENL model.

\subsubsection{E-tec Algorithm}
\label{sec:etec}

One can estimate the topological entropy of the flow without having detailed knowledge of the underlying flow using a finite number of tracer particle trajectories \cite{Thiffeault2010}. In the literature, several algorithms take a discrete set of tracer trajectories as input to estimate the topological entropy \cite{Budisic2015, Roberts2019}. The underlying principle of these algorithms is that they treat a discrete set of trajectories as braids in space-time. The maximum exponential rate at which a braid stretches material curves (represented as topological information) over a finite time period is the topological entropy of the braid. In our study, we use a doubly periodic implementation of the E-tec (Ensemble-based topological entropy calculation) algorithm~\cite{Roberts2019, EtecCode}. E-tec uses trajectories to create a triangular mesh via Delaunay triangulation, which is updated as these trajectories braid around each other. By considering material curves as topological objects within this mesh, E-tec calculates the total weight (a measure of the curve length) at each time step. A lower bound on the true topological entropy of the flow can then be estimated from the slope of the semi-log plot of weight vs. time.

\begin{figure}
 \centering
\includegraphics[width = 1.0 \columnwidth]{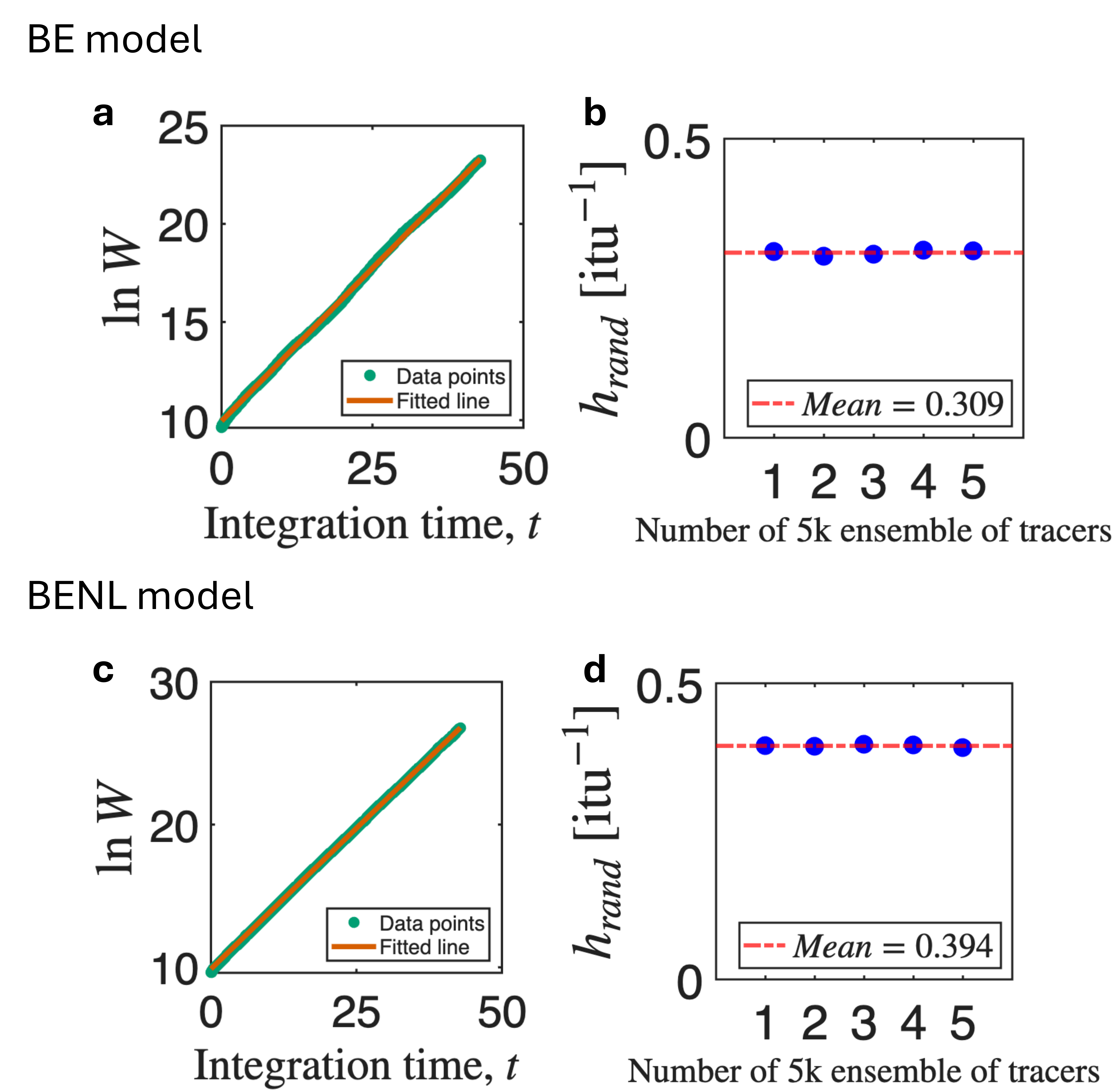}
\caption{Summary of the analysis using random tracer trajectories via the E-tec algorithm. (a,c) Semi-log plot of weight $W$ vs integration time $t$. Here, $W$ is a measure of stretching and folding of a geometric (material) curve produced by the collective motion of the passive tracers. (b,d) Blue circles are the topological entropy $h_{rand}$ calculated using the E-tec algorithm for five ensembles of $5000$ randomly initialized passive tracers. The red line is the average $h_{rand}$ over these five ensembles.  \label{fig:Randetec} }
\end{figure}

\begin{figure}
 \centering
\includegraphics[width = 1.0 \columnwidth]{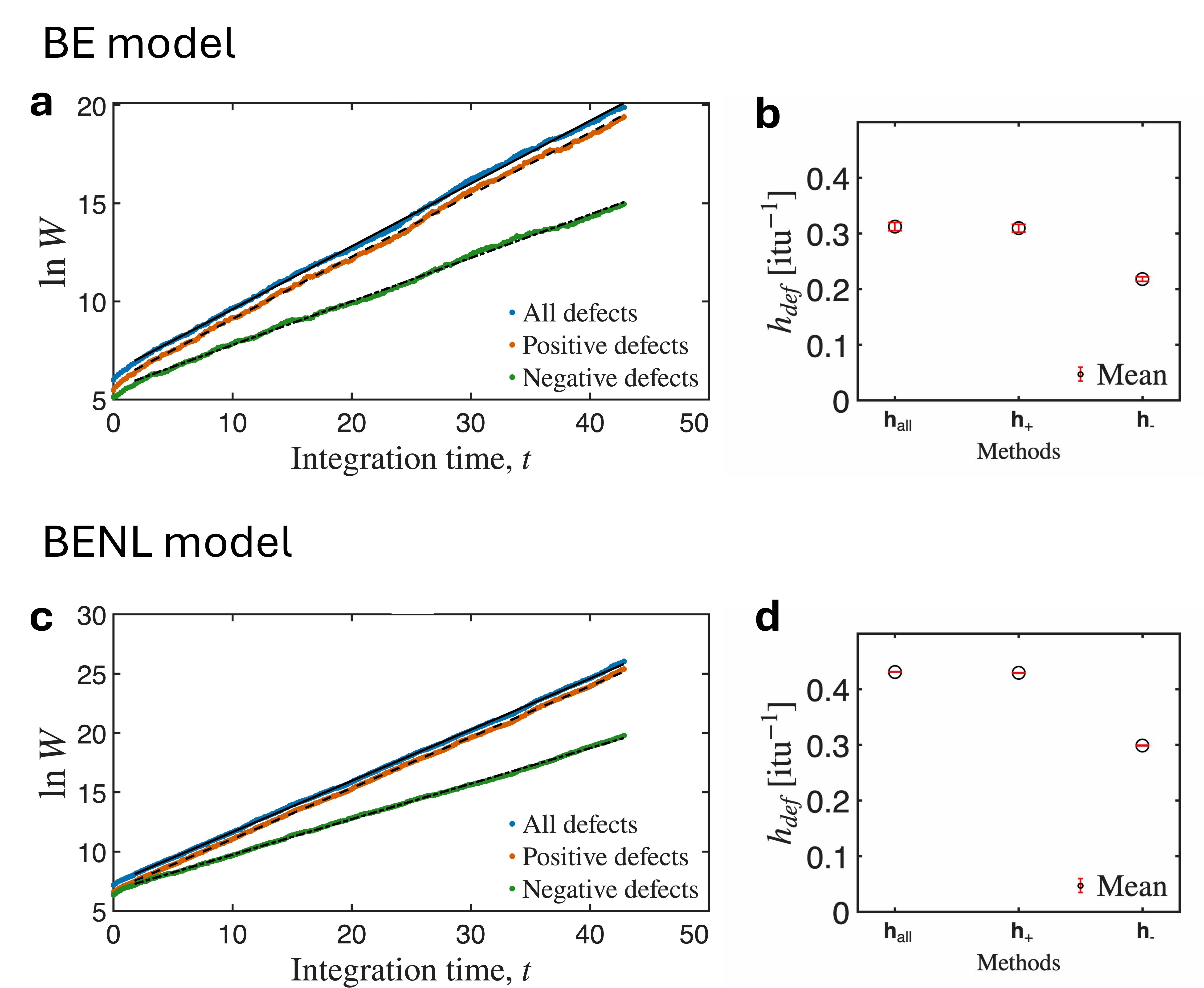}
\caption{ Summary of the E-tec analysis using topological defect trajectories. (a, c) Semi-log plot of weight $W$ vs integration time $t$. Each line in the semi-log plot represents E-tec analysis using all defect trajectories (blue), only $+1/2$ defect trajectories (yellow), and only $-1/2$ defect trajectories (green). The fitted lines are shown in black with different line styles for both the BE and BENL models. (b, d) Each circle represents the mean topological entropy $h_{def}$ per integration time for each defect trajectory group, along with the standard error of the mean across three example runs. \label{fig:defAna} }
\end{figure}


Next, we calculate the topological entropy using topological defect trajectories. E-tec requires a complete trajectory over the entire integration time.  However, topological defects are created and annihilated during that period. Therefore, to use E-tec with topological defect trajectories, we must extend these trajectories over time as passive tracers in the fluid domain. When a defect annihilates with a defect of opposite topological charge, one can imagine placing a passive tracer at that position in the fluid domain. Then we use that position as an initial condition and integrate the underlying flow field forward in time to get the trajectory until the end of the total integration time. Similarly, when defects are created at a time step, we integrate their trajectories backward in time to the initial time. Then we get entangled braids that consist of complete defect trajectories and use E-tec to calculate the topological entropy from the braiding dynamics of these trajectories. 

 To be consistent with the analysis using random trajectories, we use the same total integration time, $T = 42.86$, for the defect trajectories. Fig.~\ref{fig:defAna} summarizes the E-tec analysis using defect trajectories for both the BE and BENL models. First, we show an example of a semi-log plot of weight $W$ vs time $t$ by grouping the trajectories into three categories: all defect trajectories, only positive defect trajectories, and only negative defect trajectories (Fig.~\ref{fig:defAna}a, Fig.~\ref{fig:defAna}c). The linear fits to the semi-log plots in Fig.~\ref{fig:defAna}a yield slopes of $0.320$, $0.317$, and $0.222$ for the all-defects, positive-defects, and negative-defects trajectory sets of the BE model, respectively. The corresponding slopes in Fig.~\ref{fig:defAna}c for the BENL model are $0.431$, $0.429$, and $0.299$. These slope values represent the topological entropy in units of $\text{itu}^{-1}$. For both models, the topological entropy for positive defects is approximately the same when all defect trajectories are considered. This observation reveals that positive ($+1/2$) defects are solely responsible for the topological entropy of the flow, i.e., mixing of the fluid. In contrast, using only negative ($-1/2$) trajectories, we get considerably lower topological entropy in both models, suggesting that the contribution of negative ($-1/2$) defects to mixing is negligible. These measurements align well with the experimental observation in Ref.~\onlinecite{Tan2019}, which is reproducible in both BE and BENL models. 
 
 In Fig.~\ref{fig:defAna}b and Fig.~\ref{fig:defAna}d, we repeat the above analysis procedure for all three sets (all, positive and negative defects) of trajectories for three ensemble of the total integration time $T = 42.86$ to get an average value of the topological entropy and the standard error of their mean for both BE and BENL models (also see supplementary Fig. 3). For the BE model, we get an average $h_{all} = 0.312 \pm 0.008$ for all defects, $h_{+} = 0.309 \pm 0.007$ for positive defects and $h_{-} = 0.218 \pm 0.004$ for negative defects. For the BENL model, we get an average $h_{all} = 0.431 \pm 0.0002$ for all defects, $h_{+} = 0.429 \pm 0.0003$ for positive defects and $h_{-} = 0.299 \pm 0.001$ for negative defects. 

\subsubsection{Summary of the Topological Entropy Analysis}

\label{subsec:topEn}

\begin{figure}
 \centering
\includegraphics[width = 0.85  \columnwidth]{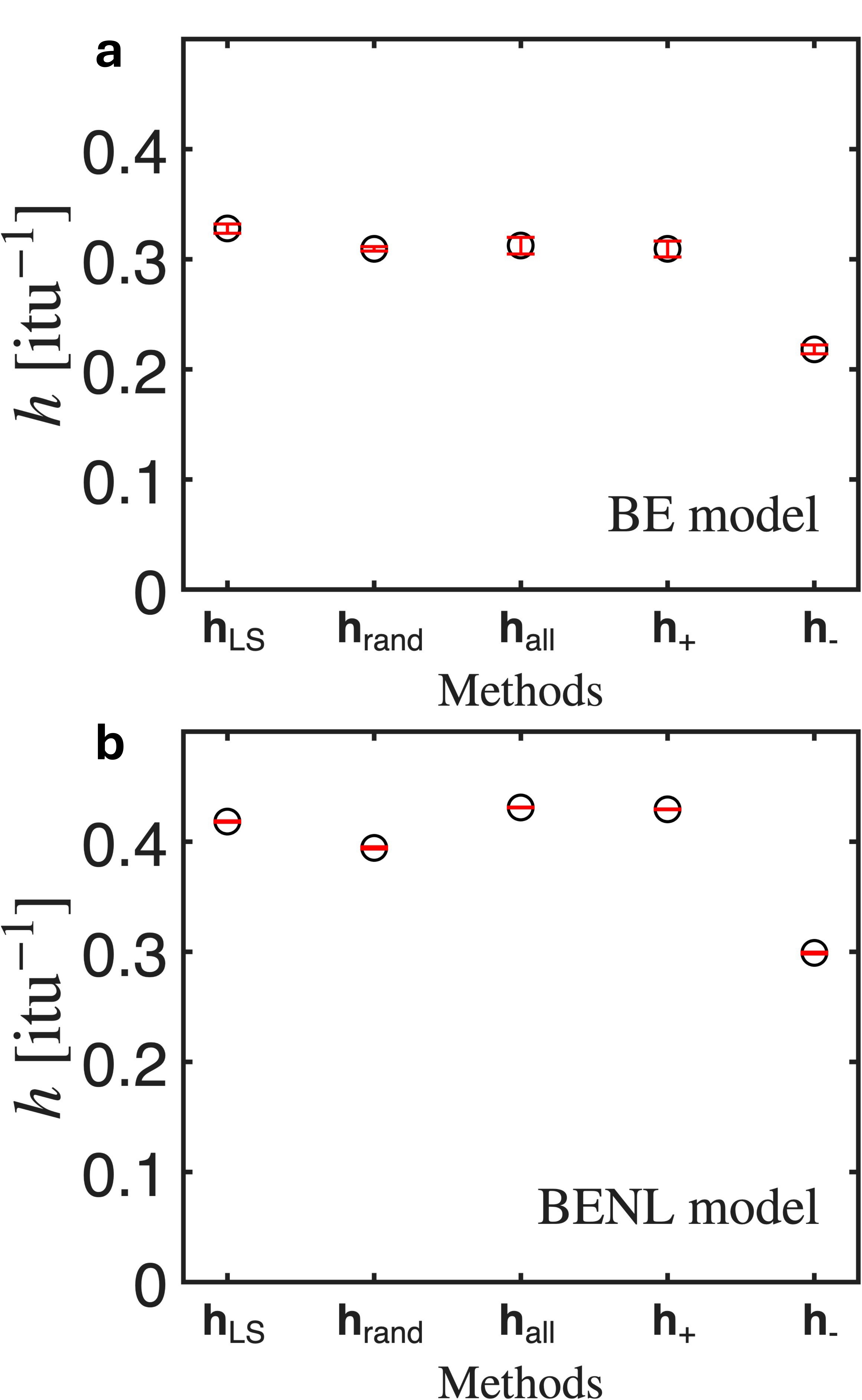}
\caption{Summary of the topological entropy analysis for the (a) BE model and the (b) BENL models. The circles are the estimates of the topological entropy of the flow, and the error bars are the standard error of the mean for each method. Topological entropy is measured in inverse integration time units. \label{fig:TEsum} }
\end{figure}

\begin{table}[htbp]
    \centering
    
    \begin{ruledtabular}
        \begin{tabular}{lcc}
            Method & BE $[itu^{-1}]$ & BENL $[{itu}^{-1}]$ \\
            \colrule
            $h_{\mathrm{LS}}$
                & $0.328 \pm 0.00420$ & $0.418 \pm 0.000604$ \\
            $h_{\mathrm{rand}}$
                & $0.309 \pm 0.00196$ & $0.394 \pm 0.000972$ \\
            $h_{\mathrm{all}}$
                & $0.312 \pm 0.00762$ & $0.431 \pm 0.000207$ \\
            $h_{+}$
                & $0.309 \pm 0.00724$ & $0.429 \pm 0.000326$ \\
             $h_{-}$
                & $0.218 \pm 0.00391$ & $0.299 \pm 0.000839$ \\
        \end{tabular}
    \end{ruledtabular}
    \caption{\label{tab:topEn_measures}
    Topological entropy measures for the BE and the BENL models. Reported numerical values are the mean and the standard error of the mean.}
\end{table}

Fig.~\ref{fig:TEsum} shows all the different methods we used to estimate the topological entropy of the flow. We also list the numerical values of the topological entropy for all methods in Table~\ref{tab:topEn_measures} for both the BE and BENL models. We note that the estimated topological entropies of  $h_{LS}$, $h_{rand}$, $h_{all}$, and $h_{+}$ are nearly equal for both the BE and BENL models.In contrast, the estimated topological entropy using negative defect trajectories $h_{-}$ is smaller than that of the other methods. Although the numerical values of $h_{+}$ and $h_{-}$ are not the same for both models, the ratio of $\frac{h_{+}}{h_{-}}$ is essentially the same (the ratio is $1.417$ for the BE model and $1.435$ for the BENL model). This suggests that the topological measures of mixing in both models capture the same information.

\subsection{Lyapunov Exponent}


The Lyapunov exponents, another fundamental measure of dynamical systems, are asymptotic measures that characterize the rates of divergence and convergence of nearby trajectories in phase space~\cite{Tancredi2001, Geist1990}. A signature of a chaotic dynamical system is that nearby trajectories diverge exponentially with time. In the context of flows, Lyapunov exponents measure the stretching of the fluid. Any fluid flow is considered chaotic if there is an emergence of at least one positive Lyapunov exponent~\cite{LU2005}. The positive Lyapunov exponent of a flow reflects the inverse time scale on which the dynamics of fluid flow become unpredictable, i.e., the extent of mixing~\cite{Aurell1996, Kleinfelter2005}.

In the literature, there are numerous methods to calculate the largest Lyapunov exponent, as well as the full spectrum of Lyapunov exponents of the flow \cite{Geist1990, sandri1996, Dieci1997, Tancredi2001, Skokos2010}. In this article, we use four numerical approaches to estimate the largest Lyapunov exponent of the flow. First, we use methods based on two nearby trajectories. Then we use the discrete QR reorthogonalization method to estimate the full Lyapunov spectrum through the evolution of orthogonal vectors in the tangent space. Lastly, we calculate the Lyapunov exponent using a director-conditioned measure of stretching specific to active nematics that requires the \emph{nematic locking principle} \cite{Tan2019}. In the following subsections, we introduce each method and then estimate the largest Lyapunov exponent of the flow.

\subsubsection{Finite Time Renormalization (FTR) Method}

The two-trajectory method using finite time renormalization (FTR) is one of the most direct ways to estimate the largest Lyapunov exponent \cite{Benettin1976, benettin1980}. The idea is to evolve two nearby passive tracers,
$\mathbf{x_{1}}(t)$ and $\mathbf{x_{2}}(t)$,
with a small initial separation distance
\begin{equation}
\delta_0 = \|\mathbf{x_{2}}(t_{0}) - \mathbf{x_1}(t_{0})\|
\end{equation}
where we consider $\mathbf{x_{1}}(t)$ to be the reference trajectory and $\mathbf{x_{2}}(t)$ the perturbed trajectory. Then we track the growth of the perturbation vector between the two trajectories over a finite-time window $t$. Then we estimate the Lyapunov exponent for that particular time window as
\begin{equation}
\lambda_t = \frac{1}{t}\ln\left(\frac{\delta_{t}}{\delta_0}\right).
\end{equation}
In practice, the separation distance cannot be allowed to grow indefinitely, since once it becomes too large, the dynamics leave the linear regime. For this reason, the perturbation vector must be renormalized to a smaller length repeatedly. However, it is also important to be cautious if the separation distance is too small relative to the simulation grid scale, as the roundoff and interpolation errors may dominate. After evolving the pair of trajectories for a fixed time $\Delta t_{\mathrm{renorm}}$, the separation distance is measured to calculate the logarithmic growth. Then we reset the perturbed trajectory at a distance $\delta_{0}$ from the reference trajectory along the current separation direction. We repeat the renormalization procedure $N$ times over the total integration time $T$ along the reference trajectory $\mathbf{x_{1}}(t)$, which allows us to estimate the largest Lyapunov exponent as
\begin{equation}
 \lambda_{FTR} = \frac{1}{T}
\sum_{k=1}^{N}
\ln\left(\frac{\delta_k}{\delta_0}\right)  \approx \lambda_{max},
\end{equation}
where $\delta_k$ is the separation at the end of the $k$th interval and $T = N \Delta t_{\mathrm{renorm}}$ is the total integration time.

\begin{figure}
 \centering
\includegraphics[width = 0.95 \columnwidth]{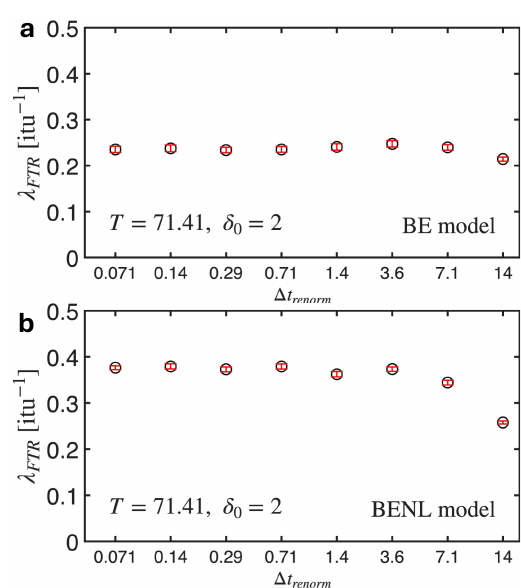}
\caption{ Summary of the analysis using the FTR method for both the BE and the BENL models. The estimate of the largest Lyapunov exponent for several renormalization time intervals $\Delta t_{renorm}$ is shown with an initial separation distance $\delta_{0} =2$. The circle is the mean over $50$ reference trajectories and the error bar is the standard error of the mean.}
\label{fig:2traj}
\end{figure}

Fig.~\ref{fig:2traj} shows the estimate of the largest Lyapunov exponent using several renormalization time intervals $\Delta t_{renorm}$ keeping the same total integration time $T = 71.41$ and the initial separation distance $\delta_{0} = 2$ for both the BE and BENL models. The estimate of the largest Lyapunov exponent we get is consistent over several short renormalization time intervals. However, the FTR method loses accuracy when we choose a longer renormalization time interval to estimate the largest Lyapunov exponent. We also used several initial separation distances $\delta_{0}$ to check the numerical stability of our implementation (supplementary Fig. S4).

\subsubsection{Finite Size Renormalization (FSR) Method}

\begin{figure}
 \centering
\includegraphics[width = 0.95 \columnwidth]{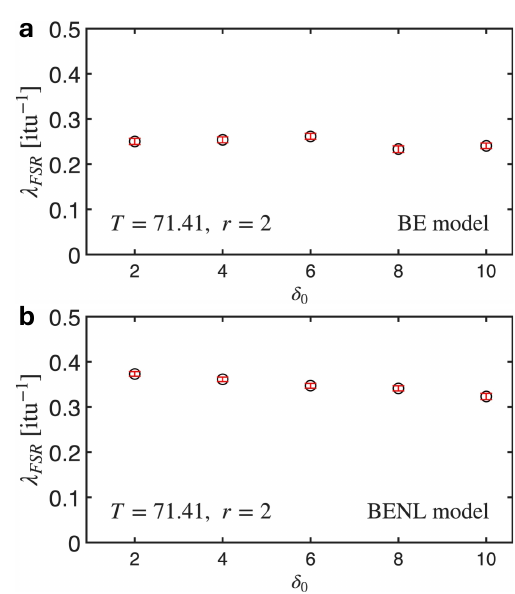}
\caption{Summary of the analysis using the FSR method for both the BE and the BENL models. The estimate of the largest Lyapunov exponent for several initial separation distances $\delta_{0}$ is shown using a \emph{growth factor} $r =2$ between nearby passive tracer trajectories. The circle represents the mean across $50$ reference trajectories, and the error bar represents the standard error of the mean.}
\label{fig:FSLE}
\end{figure}

The two-trajectory method using finite size renormalization (FSR) is another measure of chaoticity that approximates the Lyapunov exponent~\cite{Aurell1996, Aurell1997, Boffetta1998}. In the FTR method, a fixed time interval is used for the initial separation distance to grow and estimate the Lyapunov exponent of the flow. However, in the FSR method, we choose an initial separation distance $\delta_0$ between two trajectories to grow by a fixed factor $r>1$. If $\tau_f(\delta_0)$ denotes the first passage time for the initial separation distance between two passive tracers to grow from $\delta_0$ to $r\delta_0$, then the Lyapunov exponent is defined as
\begin{align}
\lambda_{FSR}(\delta_0) 
&= \left\langle \frac{1}{\tau_f(\delta_0)} \right\rangle_{\mathbf{x}(t)} \ln r  \label{eq:fsle}
\end{align}
where $\langle \cdot \rangle_{\mathbf{x}(t)}$ denotes the time-weighted average over the first passage times along the reference trajectory $\mathbf{x}(t)$. Then we collect the first passage times for an ensemble of trajectories with separation distance $\delta_0$ to get an average of first passage times. In the limit of very small separation distance, the FSR converges to the largest Lyapunov exponent \cite{Cencini2013}.
\begin{equation}
\lim_{\delta_0 \to 0} \lambda_{FSR}(\delta_0) \approx \lambda_{max}.
\end{equation}

To estimate the largest Lyapunov exponent using the FSR method, we first define a series of \emph{threshold} factors $r$. Then, for each $r$, we initialize two nearby passive tracers with a separation distance $\delta_0$ and collect the first passage times such that $\delta_0 \geq r\delta_0$ along the reference trajectory. We reset the perturbed trajectory after each collection of first passage times $\tau(\delta_f)$ by choosing a \emph{new} perturbed trajectory with the same initial distance $\delta_0$ from the reference trajectory along the current separation direction, i.e., the same method we use to initialize the perturbation vector for the FTR method. Finally, we repeat this procedure for many initial conditions, i.e., reference trajectories, to obtain an ensemble average of the first passage times to calculate an estimate of the largest Lyapunov exponent of the flow.

Fig.~\ref{fig:FSLE} shows the estimate of the largest Lyapunov exponent using the FSR method for both the BE and BENL models. We use multiple initial separation distances $\delta_0$ and collect first passage times when this separation distance doubles, i.e., the initial perturbation grows up to a factor $r=2$. Then we calculate the estimate of the largest Lyapunov exponent using Eq.~(\ref{eq:fsle}) for $50$ reference trajectories for a total integration time $T = 71.41$. We calculate the average over the number of trajectories at each separation distance $\delta_0$ to get the estimate of the largest Lyapunov exponent of the flow. We observe that the estimate of the largest Lyapunov exponent using the FSR method is fairly consistent over initial separation distances. However, the FSR converges to the maximum Lyapunov exponent of the flow when the initial separation distance is infinitesimally small. So, to get a better estimate of the largest Lyapunov exponent, smaller values of the initial separation distance $\delta_0$ are preferred. We also assessed the numerical stability of the FSR method over multiple factors $r$ to test the accuracy of our numerical implementation (supplementary Fig. S5).

\subsubsection{QR Orthonormalization Method}

The two-trajectory-based methods estimate only the largest Lyapunov exponent of the flow, since both the FTR and FSR methods use a single perturbation vector to estimate the maximum Lyapunov exponent. A natural generalization for $n$-dimensional flows is to use $n$ orthogonal perturbation vectors in the tangent space to calculate the full spectrum of Lyapunov exponents. The standard approach to get the full spectrum for a continuous-time dynamical system is to use the discrete QR orthonormalization method~\cite{Eckmann1985, Shimada1979, Benettin1980part2}. In this method, in addition to integrating the velocity $\mathbf{u}$ to get reference trajectories,  
\begin{align}
\dot{\mathbf{x}} &= \mathbf{u}(\mathbf{x}(t),t), \label{tracer}
\end{align}
we also integrate orthogonal tangent (perturbation) vectors along the reference trajectory. Let $\mathbf{\Phi}_t(\mathbf{x}_0)$ denote the flow map of a passive tracer at initial time $ t_0$ with initial position $\mathbf{x}_0$ such that
\begin{align}
    \mathbf{x}(t) = \mathbf{\Phi}_t(\mathbf{x}_0),
    \qquad \mathbf{\Phi}_{t_0}(\mathbf{x}_0) = \mathbf{x}_0 ,
\end{align}
where $\mathbf{x}(t)$ is the reference trajectory. Then the gradient of the flow map $ \nabla\mathbf{\Phi}_t(\mathbf{x}_0)$, i.e., the deformation gradient satisfies the following differential equation
 \begin{align}
        \nabla \mathbf{\dot{\Phi}_{t}}(\mathbf{x}_0)
= \nabla \mathbf{u}|_{\mathbf{x}(t)} \nabla \mathbf{\Phi_t}(\mathbf{x}_0),\label{deformGrad}
\end{align}
where $\nabla \mathbf{u}$ is the gradient of the velocity $\mathbf{u}$ calculated along the reference trajectory $\mathbf{x}(t)$. Eq.~(\ref{deformGrad}) is a matrix-valued time-dependent linear differential equation with an initial condition at time $t_0$ for flows in two dimensions, $\nabla \mathbf{\Phi}_{t_{0}}(\mathbf{x}_0) = \mathsf{I}_{2 \times 2}$, where $\mathsf{I}$ is the identity matrix. 

\begin{figure}
 \centering
\includegraphics[width = 0.95 \columnwidth]{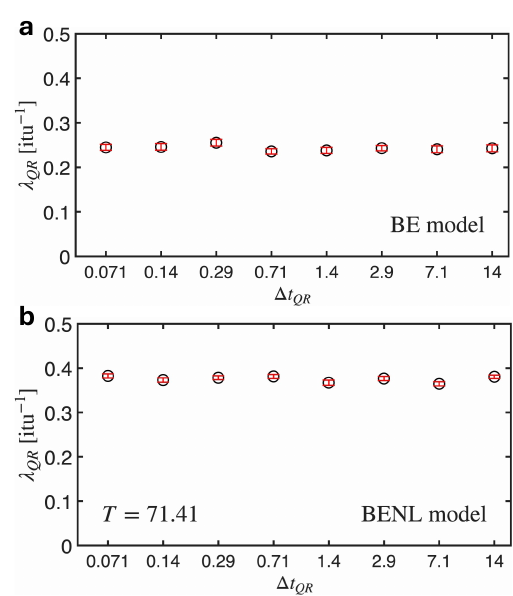}
\caption{Summary of the analysis using the QR orthonormalization method for both the BE and the BENL models. The estimate of the largest Lyapunov exponent for several orthogonalization time intervals $\Delta t_{QR}$ is shown. The circle is the mean over $50$ reference trajectories and the error bar is the standard error of the mean.}
\label{fig:Qrall}
\end{figure}

We numerically integrate Eq.~(\ref{tracer}) and Eq.~(\ref{deformGrad}) simultaneously to obtain the solution, $\mathsf{Y} = \nabla \mathbf{\Phi_{t}}$, of the differential Eq.~(\ref{deformGrad}). The columns of $\mathsf{Y}$ represent the evolution of the initial orthogonal tangent (perturbation) vectors. Over the integration time, these perturbation vectors tend to align with the eigenvector associated with the largest eigenvalue, i.e., the direction of largest stretching (most unstable direction), thereby causing a loss of linear independence among the perturbation vectors and making the estimation of the Lyapunov exponents erroneous. To solve this issue, we orthonormalize the perturbation vectors over a short (discrete) time interval $\Delta t_{\mathrm{QR}}$ and represent $\mathsf{Y}$ by its QR-decomposition $\mathsf{Q}\mathsf{R}$,
where $\mathsf{Q}$ is an orthonormal matrix and $\mathsf{R}$ is an upper triangular matrix. Columns of $\mathsf{Q}$ represent the evolved set of perturbation vectors and the diagonal elements of $\mathsf{R}$ contain the local stretching factors for each time interval $\Delta t_{\mathrm{QR}}$. We repeat this procedure along the reference trajectory $N$ times to get an estimate of the Lyapunov exponents as time averages of the logarithms of diagonal elements of the upper triangular matrix $\mathsf{R}_k$:
\begin{equation}
\lambda_i =
\frac{1}{T}
\sum_{k=1}^{N}
\ln \left| (\mathsf{R}_k)_{ii} \right|,
\quad i=1,2.
\end{equation}
where $T= N\Delta t_{\mathrm{QR}}$ is the total integration time and where the diagonal elements of $\mathsf{R}_k$ are ordered $|(\mathsf{R}_k)_{11}| \ge |(\mathsf{R}_k)_{22}|$.

\begin{figure}
 \centering
\includegraphics[width = 0.95 \columnwidth]{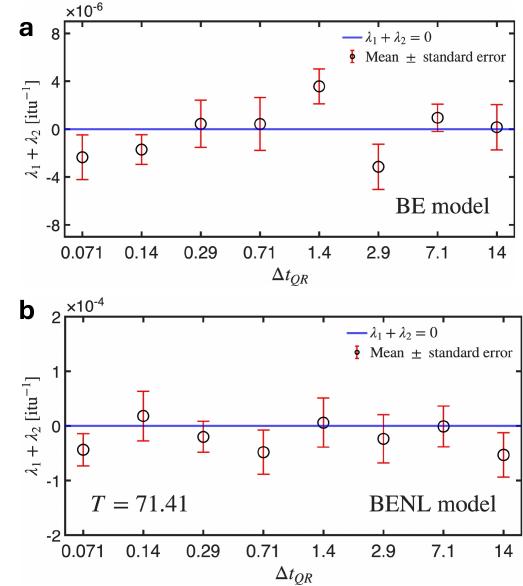}
\caption{The estimate of the sum of the Lyapunov exponents $\lambda_1 +\lambda_2$  for both the BE and the BENL models using the QR orthonormalization method. The blue horizontal line $\lambda_1 +\lambda_2 = 0$ is the expected sum for a two-dimensional incompressible fluid. The circles are the mean of the sum of Lyapunov exponents over $50$ reference trajectories, and the error bar is the standard error of the mean.}
\label{fig:SumLEQR}
\end{figure}

Fig.~\ref{fig:Qrall} shows the summary of the analysis of estimating the largest Lyapunov exponent for both models using $50$ randomly chosen reference trajectories over a total integration time $T = 71.41$. The estimate is consistent over the time period of QR decomposition $\Delta t_{QR}$. As both models (BE and BENL) treat microtubule-based active nematic flow as a 2D incompressible fluid, the sum of the Lyapunov exponents $\lambda_1 + \lambda_2$ must be zero to conserve the area of phase space. We check this criterion in Fig.~\ref{fig:SumLEQR} by plotting the sum of Lyapunov exponents $\lambda_1 +\lambda_2$ for each time interval $\Delta t_{QR}$ and get approximately zero for both models.

\subsubsection{Using Director-Aligned Perturbation Vector}

 All previous methods for calculating the largest Lyapunov exponent rely on identifying the eigenvector of maximum stretching and its eigenvalue along the reference trajectory. The FTR and FSR methods both rely on a single perturbation vector, while the QR orthonormalization method relies on an orthogonal basis of perturbation vectors to calculate an estimate of the largest Lyapunov exponent. Ref.~\onlinecite{Tan2019} calculated an estimate from the stretching along the director using experimental data for microtubule-based active nematic fluids.  Ref.~\onlinecite{Tan2019} reported that their estimated Lyapunov exponent is slightly smaller than the topological entropy. The apparent small difference between topological entropy measures and the Lyapunov exponent in the experiment infers that the calculated Lyapunov exponent along the director is indeed an estimate of the largest Lyapunov exponent of the flow. So, instead of searching for eigenvectors that maximize stretching along any reference trajectory in the fluid domain, one can choose a perturbation vector along the director and measure the extent of stretching.

 To calculate an estimate of the Lyapunov exponent of the microtubule-based active nematic fluid flow along the direction of the director, Ref.~\onlinecite{Tan2019} calculated the following
 \begin{equation}
\hat{\mathbf n} \cdot \big ( \nabla\mathbf u\ \big ) \hat{\mathbf n}, 
\label{eq:quant}
\end{equation}
\noindent where $\hat{\mathbf n}$ is the unit director and $\mathbf{u}$ is the fluid velocity. We know from the kinematics of the evolution of the material line that the fractional growth rate of an infinitesimal line element $\delta l$ oriented along a unit vector $\hat{\mathbf n}$ satisfies \cite{Cartwright1999}
\begin{equation}
\frac{d}{dt}
\ln \delta l
= \hat{\mathbf n} \cdot \mathbf E \hat{\mathbf n}, 
\end{equation}
where $\mathbf{E}$ is the symmetric strain-rate tensor.  Since only the symmetric part of the velocity gradient contributes to the local stretching of the fluid, the quantity in Eq.~(\ref{eq:quant}) measures the instantaneous stretching rate experienced by the fluid along the director. Under ergodic conditions, Ref.~\onlinecite{Tan2019} calculated an estimate of the largest Lyapunov exponent
\begin{equation}
\lambda_\mathbf n = \Big \langle \hat{\mathbf n} \cdot \nabla \mathbf u\, \hat{\mathbf n} \Big \rangle_{(\mathbf r,t)}, 
\label{eq:lambdan}
\end{equation}
where $( \mathbf r,t)$ represents the average in the spatial and time domains. Eq.~(\ref{eq:lambdan}) 
measures the long-time average of extensile material stretching (deformation) along the director. The details of the derivation of Eq.~(\ref{eq:lambdan}) can be found in Ref.~\onlinecite{Tan2019}. We can also rewrite Eq.~(\ref{eq:lambdan}) in terms of the $\mathsf{Q}$-tensor,
\begin{equation}
\lambda_\mathbf{n}
= \Big \langle \frac{1}{S}\,\mathrm{Tr}\!\left(\mathsf{Q}\,\nabla\mathbf{u}\right) \Big \rangle_{(\mathbf r,t)}, \label{eq:lambdaQfull}
\end{equation}
where $S$ is the scalar order parameter. Due to steric constraints in a microtubule-based active nematic fluid, $S \approx 1$ nearly everywhere except in small, localized regions of high curvature, i.e., in the vicinity of topological defects. Then Eq.~(\ref{eq:lambdaQfull}) becomes
\begin{equation}
\lambda_\mathsf{Q}
= \Big \langle \mathrm{Tr}\!\left(\mathsf{Q}\,\nabla\mathbf{u}\right) \Big \rangle_{(\mathbf r,t)} \approx \lambda_\mathbf n. \label{eq:lambdaQ}
\end{equation}

\begin{figure}
 \centering
\includegraphics[width = 0.85 \columnwidth]{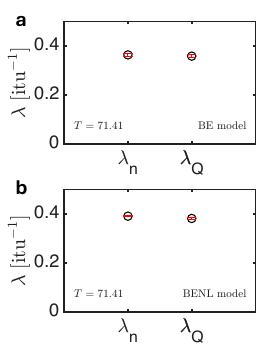}
\caption{Summary of the analysis for both the a) BE and b) BENL models using Eq.~(\ref{eq:lambdan}) and Eq.~(\ref{eq:lambdaQ}). The estimates of the largest Lyapunov exponent for a total integration time $T=71.41$ are shown.  The error bar is the correlation-corrected standard error of the mean.}
\label{fig:LEdfplots}
\end{figure}

 To estimate the Lyapunov exponent, we first calculate the mean by taking spatial averages of the quantities in Eq.~(\ref{eq:lambdan}) and Eq.~(\ref{eq:lambdaQ}) at each time step, thereby constructing a time series. Then we take the average of the series to estimate the Lyapunov exponent, as shown in Fig.~\ref{fig:LEdfplots} for both the BE and BENL models. To compute error bars, we note that the time series is correlated, so we must determine the number of independent measures over the total integration time $T$. We follow the \emph{blocking} method described in Ref.~\onlinecite{Flyvbjerg1998} to get a correlation-corrected standard error of the mean.

\subsubsection{Summary of the Lyapunov Exponent Analysis}

\begin{figure}
 \centering
\includegraphics[width = 0.95 \columnwidth]{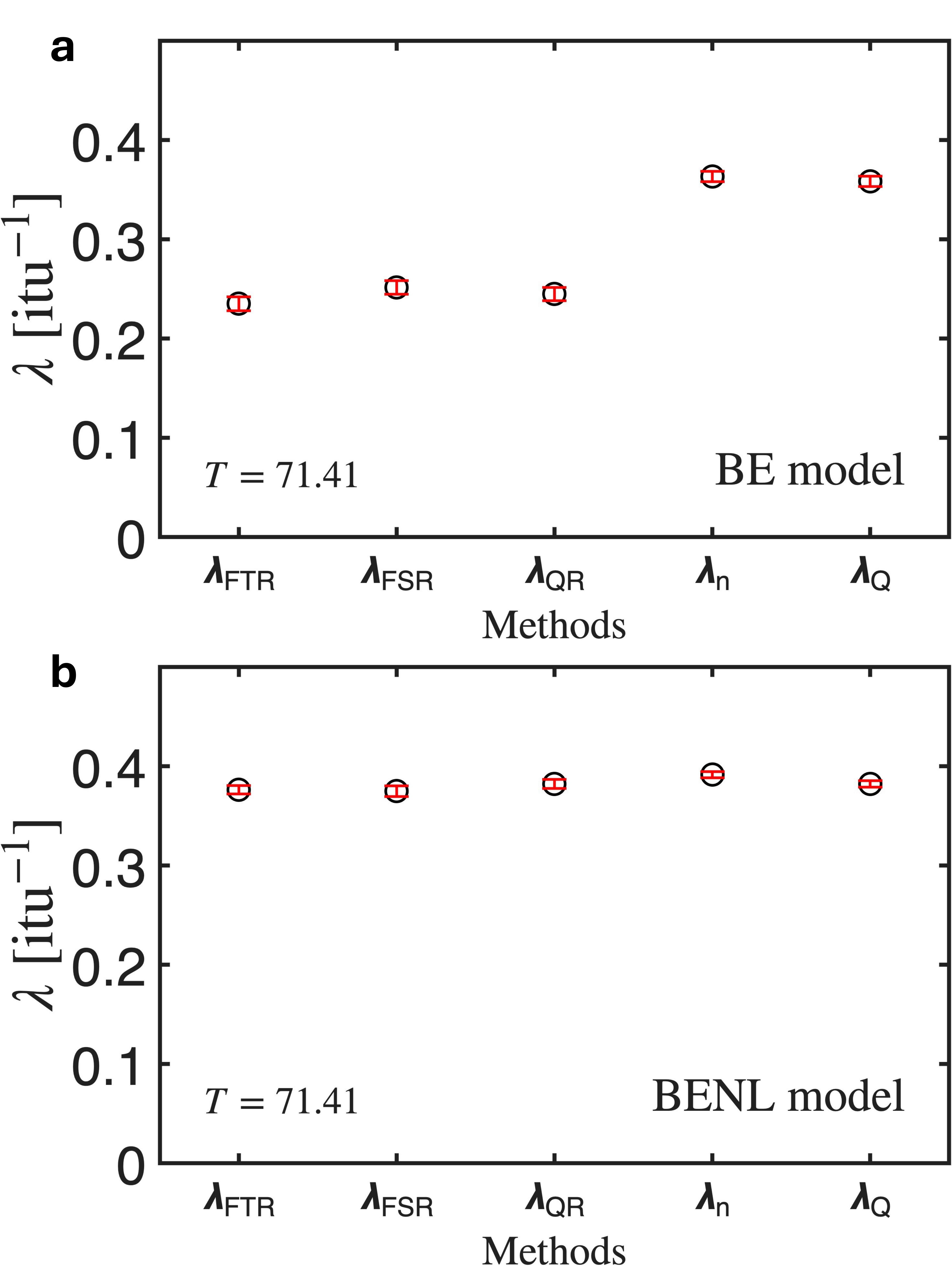}
\caption{Summary of the Lyapunov exponent analysis for the (a) BE model and the (b) BENL model. The circles are the estimates of the largest Lyapunov exponent and the error bars are the standard error of the mean for each method. The units for the Lyapunov exponent are in inverse integration time. }
 \label{fig:LEsum}
\end{figure}

We summarize the analysis for all methods to estimate the Lyapunov exponent $\lambda$ in Fig.~\ref {fig:LEsum} and list the numerical values of $\lambda$ for all methods in Table~\ref{tab:LE_measures} for both the BE model and the BENL model. There are several parameters for each method, e.g., initial separation distance $\delta_0$, time interval $\Delta t$ to renormalize (for the two-trajectory-based methods) or orthogonalize (for the QR method) the perturbation vector, and the threshold factor $r$ specifically for the FSR method. So, for the summary plot in Fig.~\ref{fig:LEsum}, we aim to be consistent with the parameters used across all methods while respecting each method's limitations. For that reason, we choose the initial separation distance $\delta_0 = 2$ for both the two-trajectory-based methods (FTR and FSR). We use the integration time interval $\Delta t = 0.07$ for both the FTR and the QR orthogonalization methods for renormalization and orthogonalization, respectively. We use the same total integration time $T = 71.41$ as in other methods for the director-field-conditioned measure of the Lyapunov exponent. Lastly, we use the threshold factor $r=2$ to obtain the first-passage times and estimate the largest Lyapunov exponent using the FSR method.

\label{subsec:LE}

\begin{table}[htbp]
\centering
\begin{ruledtabular}
    \begin{tabular}{lcc}
        Method & BE $[{itu}^{-1}]$ & BENL $[{itu}^{-1}]$ \\
        \colrule
        $\lambda_{\mathrm{FTR}}$
            & $0.235 \pm 0.00704$ & $0.376 \pm 0.00423$ \\
        $\lambda_{\mathrm{FSR}}$
            & $0.251 \pm 0.00680$ & $0.375 \pm 0.00522$ \\
        $\lambda_{\mathrm{QR}}$
            & $0.245 \pm 0.00657$ & $0.382 \pm 0.00468$ \\
        $\lambda_{n}$
            & $0.363 \pm 0.00528$ & $0.392 \pm 0.00319$ \\
        $\lambda_{Q}$
            & $0.358 \pm 0.00528$ & $0.382 \pm 0.00319$ \\
    \end{tabular}
\end{ruledtabular}
\caption{\label{tab:LE_measures}
Lyapunov exponent measures for the BE and the BENL models. Reported numerical values are the mean and the standard error of the mean.}
\end{table}

It is clearly evident in Fig.~\ref{fig:LEsum} that the director-aligned estimate of the Lyapunov exponent (Eq.~(\ref{eq:lambdan}) and Eq.~(\ref{eq:lambdaQ})) using the simulated flow and the orientational field of the BE model does not match the other established methods of estimating the largest Lyapunov exponent of a flow. In contrast, all measures of the Lyapunov exponent using the BENL model agree with each other within numerical error. Experiments on the microtubule-based active nematic fluid in Ref.~\onlinecite{Tan2019} report that the director-aligned perturbation vector estimate of the Lyapunov exponent is slightly smaller than the measures of the topological entropy. This implies that the estimated Lyapunov exponent measured from the experimental data using Eq.~(\ref{eq:lambdan}) is the largest Lyapunov exponent of the flow. In Fig.~\ref{fig:LEsum}b, the well-established methods of estimating the largest Lyapunov exponent using the fluid velocity match the director-aligned measure for the BENL model. This signifies that the experimentally measured Lyapunov exponent aligns better with the BENL model, thereby supporting the adoption of the \emph{nematic locking principle} to model the microtubule-based active nematic fluid.

\section{Conclusions}
\label{sec:end}

\begin{figure}
 \centering
\includegraphics[width = 0.95 \columnwidth]{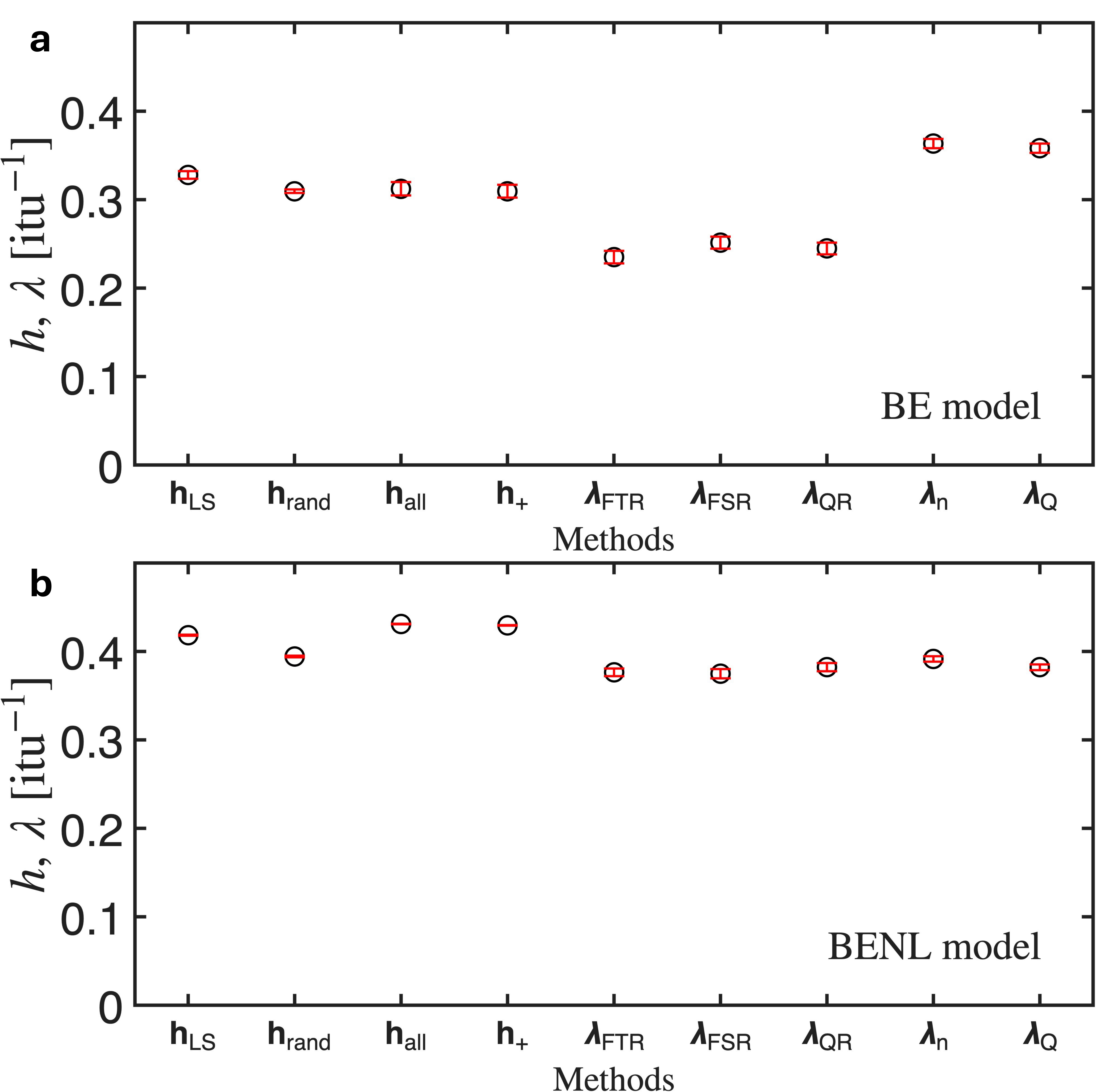}
\caption{Summary of the measures of chaotic advection for the (a) BE and the (b) BENL models. The circles are the estimates of the topological entropy $h$ and the largest Lyapunov exponent $\lambda$. The error bars are the standard error of the mean for each method.\label{fig:TEplusLEsum} }
\end{figure}

We have employed several methods to estimate the topological entropy $h$ and the Lyapunov exponent $\lambda$ for the two-dimensional microtubule-based active nematic fluid using two continuum models (BE and BENL). Fig.~\ref{fig:TEplusLEsum} shows all the measures of chaotic advection, combining the summary figures in Fig.~\ref{fig:TEsum} and Fig.~\ref{fig:LEsum}. We excluded topological entropy measured from negative-defect trajectories, as they do not encode the full topological entropy of the flow. If we set aside the Lyapunov exponent analysis using the director-aligned perturbation vector for a moment, we observe that in both models (BE and BENL), all measures of chaotic advection satisfy the well-established mathematical inequality in a two-dimensional flow $h \geq \lambda$. However, the crucial difference between the BE and BENL models lies in the estimate of the Lyapunov exponent using the director-aligned perturbation vector. For the BENL model, the Lyapunov exponent obtained using the director-aligned perturbation vector method is approximately the same as that obtained by other established methods of measuring the largest Lyapunov exponent, and different measures of topological entropy are slightly larger than all different measures of the largest Lyapunov exponent, which matches the experimental measurements of chaotic advection in Ref.~\onlinecite{Tan2019}. 

\begin{figure}
 \centering
\includegraphics[width = 0.95  \columnwidth]{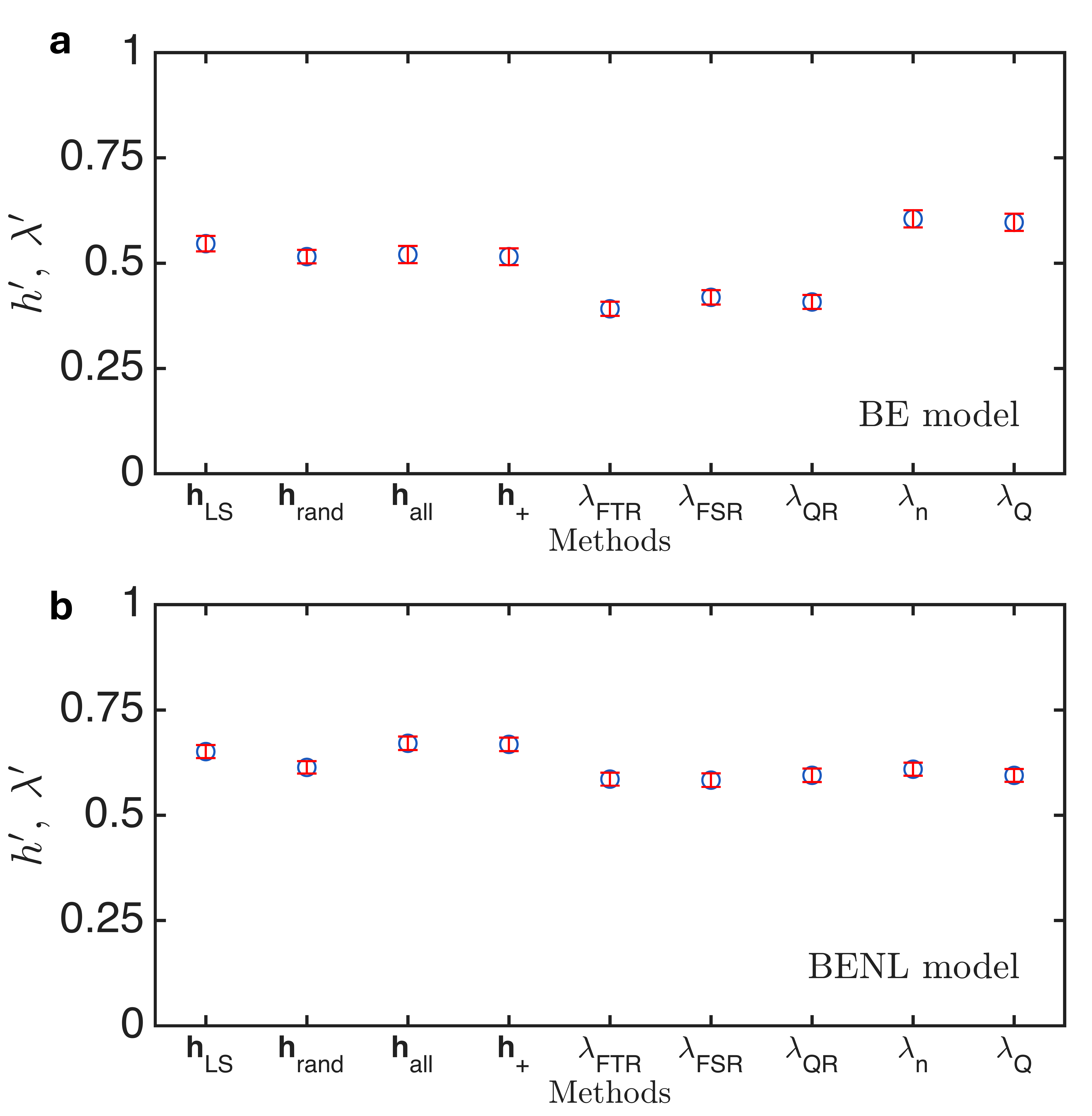}
\caption{ The dimensionless  (blue circles) of chaotic advection for the (a) BE and the (b) BENL models. Error bars are calculated using, e.g., $\Delta h' = h' \left[(\Delta h/h)^2 + (\Delta \tau /\tau)^2\right]^{1/2}$. We follow the same approach to calculate $\Delta \lambda'$. \label{fig:dimless} }
\end{figure}

We estimate chaotic advection measures using inverse integration time units, i.e., $\mathrm{itu^{-1}}$. To better compare the BE and BENL models, we measure a characteristic time scale $\tau = l_{u}/u_{rms}$ of the flow to non-dimensionalize chaotic advection measures using the same approach as in Ref.~\onlinecite{Tan2019}. The length scale $l_{u}$ is the separation at which the velocity–velocity correlation function falls to half its maximum value and $u_{rms}$ is the average root-mean-square velocity. We obtain $\tau_{BE} = 1.666 \pm 0.051$ and $\tau_{BENL} = 1.556 \pm 0.037$ in integration time units for the BE and BENL models, respectively. The reported error of the characteristic timescale $\tau$ combines error in measuring characteristic length $l_{u}$ and velocity $u_{rms}$, i.e., $\Delta \tau = \tau\left[(\Delta l_u/l_u)^2 + (\Delta u_{\mathrm{rms}}/u_{\mathrm{rms}})^2\right]^{1/2}$. We then obtain the dimensionless topological entropy $h' = \tau h$, and the dimensionless Lyapunov exponent $\lambda' = \tau \lambda$ (Fig.~\ref{fig:dimless}) and observe a similar trend in dimensionless measures of chaotic advection as in its dimensional counterpart (Fig.~\ref{fig:TEplusLEsum}).

\begin{figure}
 \centering
\includegraphics[width = \columnwidth]{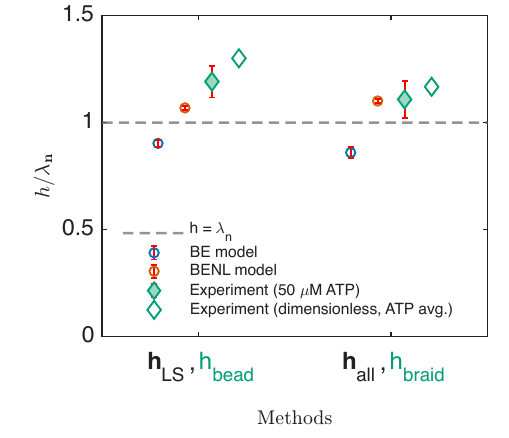}
\caption{The ratio of the topological entropy measures and the Lyapunov exponent estimated using the director-aligned perturbation vector method. We use line stretching ($\mathsf{h_{LS}}$) and topological entropy measured using all defect trajectories ($\mathsf{h_{all}}$) via the E-tec algorithm for both the BE (blue circles) and the BENL (golden circles) models. We use the analogous experimental measurement methods (green text) using the published data (green diamonds) in Ref.~\onlinecite{Tan2019}. The filled green diamonds are the ratios using the experimental data at $50\,\mu$M ATP concentration. The open green diamonds are the ratios obtained from reported average values over six ATP concentrations (numerical values of the error were not reported). Error bars on the ratio are calculated using $\Delta(h/\lambda_{\mathbf{n}}) = (h/\lambda_{\mathbf{n}}) \left[\left(\Delta h/h\right)^{2} + \left(\Delta \lambda_{\mathbf{n}}/\lambda_{\mathbf{n}}\right)^{2}\right]^{1/2}$. \label{fig:ratio}}
\end{figure}

In Fig.~\ref{fig:ratio}, we compare the chaotic advection measures from both models (BE and BENL) with experimental measurements published in Ref.~\onlinecite{Tan2019} by taking the ratio between topological entropy measures and the estimated Lyapunov exponent from the director-aligned perturbation vector method, i.e., $h/\lambda_\mathbf{n}$. The advantage of using the ratio $h/\lambda_\mathbf{n}$ is that it cancels out the time units. Therefore, the ratio $h/\lambda_\mathbf{n}$ is unaffected by the choice of unit of time and the definition of the correlation function from which we determine the characteristic timescale $\tau$. The ratio $h/\lambda_\mathbf{n}$ tests the expected 
bound $h/\lambda \geq 1 $ directly for an incompressible (area-preserving) two-dimensional flow. We observe in Fig.~\ref{fig:ratio} that the two continuum models (BE and BENL) fall on opposite sides of this bound. The ratio using the results from the BE model lies
below the bound for both entropy measures ($h_{\mathrm{LS}}/\lambda_{\mathbf{n}} = 0.903 \pm 0.018$,
$h_{\mathrm{all}}/\lambda_{\mathbf{n}} = 0.860 \pm 0.024$). In contrast, the ratio using the BENL model lies above the bound for both entropy measures
($ h_{\mathrm{LS}}/\lambda_{\mathbf{n}} = 1.069 \pm 0.009$ and $h_{\mathrm{all}}/\lambda_{\mathbf{n}} = 1.101 \pm 0.009$). To compare with the experimental measurements, we use two types (dimensional and dimensionless) of published data from analogous experimental measurement methods for $h_{\mathrm{LS}} (h_{\mathrm{bead}})$,  $h_{\mathrm{all}} (h_{\mathrm{braid}})$, and $\lambda_{\mathbf{n}}$ in Ref.~\onlinecite{Tan2019}. We first use the experimental measures reported in physical time units ($s^{-1}$) for the $50\,\mu$M ATP concentration to obtain the following ratios $h_{\mathrm{bead}}/\lambda_{\mathbf{n}} = 1.192 \pm 0.074$ and
$h_{\mathrm{braid}}/\lambda_{\mathbf{n}} = 1.108 \pm 0.087$. We also use their reported 
dimensionless values averaged over six ATP concentrations to obtain ratios $h'_{\mathrm{bead}}/\lambda'_{\mathbf{n}} = 1.300$ and $h'_{\mathrm{braid}}/\lambda'_{\mathbf{n}} = 1.167$. Note that the ratios we calculate using both continuum models (BE and BENL) are for a single active length scale $l_{a}$. However, whether we use dimensional or dimensionless experimental values to obtain $h/\lambda_\mathbf{n}$, both numerical values sit well above 1. The BENL model thus satisfies the expected ratio of $h/\lambda_\mathbf{n}$ for microtubule-based active nematic fluids, whereas the BE model does not. In addition to the arguments in Ref.~\onlinecite{Mitchell2025}, this is further evidence that the BENL model is more experimentally consistent than the BE model for the microtubule-based active nematic fluid.

Finally, it would be interesting to explore these measures of chaotic advection to analyze other types of two-dimensional active fluids, both living~\cite{Zhou2014} and synthetic~\cite{Sokolov2025}, using both simulated and experimental flow fields.

\acknowledgments

This work was financially supported by the US Department of Energy under
grant DE-SC0025803, by the NSF grant DMR-2225543, and finally by the
University of California Office of the President under grant M25PL8991
(the UC Active Matter Hub). We thank Spencer Smith for access to the updated version of the E-tec algorithm. 

\bibliographystyle{apsrev4-2}
\bibliography{chaoticmixingMS}

\end{document}